\documentclass[twocolumn, prl,showkeys]{revtex4}
\pdfoutput=1

	\usepackage[utf8]{inputenc}
	\usepackage[T1]{fontenc}
	\usepackage{braket}
	\usepackage{natbib}

	\usepackage{amsmath}
	\usepackage{makeidx}
	\usepackage{amsfonts}
	\usepackage[usenames,dvipsnames]{pstricks}
	\usepackage[colorlinks,hyperindex]{hyperref}
	\hypersetup
	{
		colorlinks,%
		citecolor=black,%
		linkcolor=black,%
		urlcolor=black,%
	}

\makeindex

\usepackage{graphicx}

\usepackage{%
}

\usepackage[super]{nth}

\usepackage{xcolor}

\renewcommand{\eqref}[1]{Eq.~(\ref{#1})}
\newcommand{\eqrefs}[2]{Eqs.~(\ref{#1}) and (\ref{#2})}
\newcommand{\eqreft}[2]{Eqs.~(\ref{#1}) -- (\ref{#2})}

\providecommand{\e}[1]{\ensuremath{\times 10^{#1}}}

\let\myrm\mathrm 
\let\mybf\textbf 

\let\savecite\cite

\newcommand{\refcite}[1]{Ref.\,\savecite{#1}}

\newcommand{\figref}[1]{Fig.\,(\ref{#1})}
\newcommand{\figrefS}[2]{Fig.\,(\ref{#1}#2)}

\newcommand{\authemph}[1]{\emph{#1}}

\newcommand{\removedD}[1]{{\color{gray}{#1}}}
\renewcommand{\removedD}[1]{{}} 

\newcommand{\done}[1]{}
\newcommand{\unit}[1]{\,\myrm{#1}}

\newcommand*{\chem}[2]{\ensuremath{\text{#1}_{#2}}}

\def\figurepath{./Figures/}

\newcommand{\makefig}[5]{
\begin{figure}[!htp] 
  \begin{center}
     \includegraphics[keepaspectratio,#2]{\figurepath /#1}
     \caption[#4]{#3}
     \label{#5} 
  \end{center}
\end{figure}
}

\newcommand{\makefigfreetop}[5]{
\begin{figure}[t] 
  \begin{center}
     \includegraphics[keepaspectratio,#2]{\figurepath /#1}
     \caption[#4]{#3}
     \label{#5} 
  \end{center}
\end{figure}
}

\begin{document}

\title{Non-Resonant Thermal Admittance Spectroscopy}
	
\author{Deniz Bozyigit}
\email{denizb@iis.ee.ethz.ch}
\affiliation{Laboratory for Nanoelectronics, Department of Information Technology and Electrical Engineering, Eidgenoessische Technische Hochschule Zurich}
\author{Vanessa Wood}
\email{vwood@ethz.ch}
\affiliation{Laboratory for Nanoelectronics, Department of Information Technology and Electrical Engineering, Eidgenoessische Technische Hochschule Zurich}
\date{\today}

\begin{abstract}
Thermal Admittance Spectroscopy (TAS) as become a popular technique to determine trap state density and energetic position in semiconductors. In the limit of a large number of trap states (>$10^{16} \unit{cm^{-3}}$), Fermi-level pinning undermines the assumptions used in the analysis of TAS data, which leads to a significant underestimation of the trap state density. Here, we develop the tools to detect and account for the occurrence of Fermi-level pinning in TAS measurements.
\end{abstract}


\maketitle

\def\figurepath{./}



Understanding the defects of semiconductors is central to making them useful in applications. Thermal Admittance Spectroscopy (TAS) was first devised for trap state characterization in Cu(In,Ga)\chem{Se}{2} solar cells by \authemph{Walter} et al.\cite{Walter1996} For polycrystalline materials, TAS has been proven to be a powerful technique that is able to resolve the energetic distribution of trap states -  in contrast to DLTS, which is most applicable in the presence of discrete trap states. Here we discuss the relevant theory underlying TAS and the experimental technique. We show typical results that are obtained for PbS NC-solids, which motivates us to extend the model with which one analyzes TAS data to account for Fermi-level pinning in the presence of a large number of trap states.

\section{Theory}

The change in occupation of a trap state ($n_\myrm{T}$) is governed by capture and emission of electrons and holes, described by the capture coefficients ($\beta_n, \beta_p$ respectively). Following the derivation in \refcite{Walter1996}, we only consider the interaction with electrons in the conduction band and neglect the presence of holes in the valence band. The differential equation for the occupation of the trap state is given by:
\begin{align}
\begin{split}
\label{eq:TASdeq}
\frac{d n_\myrm{T}}{dt} = &\beta_n N_\myrm{c} (N_\myrm{T} - n_\myrm{T})e^{-(E_\myrm{c}-E_\myrm{F})/kT} \\ %
- &\beta_n N_\myrm{c} n_\myrm{T} e^{-(E_\myrm{c}-E_\myrm{T})/kT}  
\end{split}
\end{align}
where $N_\myrm{T}$ and $E_\myrm{T}$ are the density and energy of the trap state, $N_\myrm{c}$ the effective density of states of the conduction band and $E_\myrm{F}$ the Fermi-energy. If we apply a small external voltage signal ($\tilde{u}_\myrm{ext}$) at a frequency ($\omega$), we modulate the Fermi-energy around position of the trap state ($\tilde{u}_n$). The small-signal change in the trap occupation following this change in the Fermi-energy can be calculated from \eqref{eq:TASdeq}:
\begin{align}\label{eq:TASssnt}
%
%
\tilde{n}_\myrm{T} = N_\myrm{T} f\left(\frac{E_\myrm{F} - E_\myrm{T}}{kT}\right) \frac{q \tilde{u}_n}{kT} \frac{\nu_{00}}{i\,\omega + \omega_0} e^{-(E_\myrm{c}-E_\myrm{F})/kT},
\end{align}
where $f$ the Fermi-Dirac distribution and $\omega_0$ is the resonance frequency given by 
\begin{align}\label{eq:TASomega0}
\omega_0 = &\nu _{00}T^2 e^{-\frac{E_\myrm{c} - E_\myrm{F}}{kT}} (1+e^{-\frac{E_\myrm{T} - E_\myrm{F}}{kT} }).
\end{align}
Here, we have introduced the reduced attempt-frequency ($\nu_{00}$) by
\begin{align}\label{eq:JMCCnu0rep}
\beta _{\myrm{n}}N_{\myrm{C}} = \nu _{00}T^2 = {\sigma _{\myrm{T}}}{\Gamma _n}{T^2}.
\end{align}

The small signal change in the trap occupation described by \eqref{eq:TASssnt} can be expressed in term of a capacitance:
\begin{align}
C_\myrm{T} &= \frac{-e\,\text{Im}(\tilde{n}_\myrm{T}) \,\omega_0}{\tilde{u}_\myrm{ext}\,\omega } \\
		   &= \frac{e^2 N_\myrm{T}(E_\myrm{T})}{kT} \frac{\tilde{u}_n}{\tilde{u}_\myrm{ext}} \frac{\omega_0^2}{\omega^2 + \omega_0^2} f'\left(\frac{E_\myrm{T} - E_\myrm{F}}{kT}\right).\label{eq:TASchemcap}
\end{align}

\eqref{eq:TASchemcap} is instructive to understand under which conditions a trap state can show a capacitive response. The first term ($e^2 N_\myrm{T}/kT$) is the maximal capacitance of the trap state in units of $\unit{F/cm^3}$. The second term  ($\tilde{u}_n/\tilde{u}_\myrm{ext}$) describes which fraction of the externally applied small signal potential is translated into a small signal change of the Fermi-energy at the trap state. The third term gives the frequency dependence, which corresponds to a step function around the resonance frequency ($\omega_0$) (See \figrefS{fig:fTAStheoRes}{a}). 

The last term in \eqref{eq:TASchemcap} is the first derivative of the Fermi-Dirac distribution ($f'$), which is strongly peaked around $E_\myrm{T} = E_\myrm{F}$. Due to the sampling character of $f'$, all capacitive contributions for traps away from the Fermi-energy ($E_\myrm{T} \neq E_\myrm{F}$) are neglected in the next step of the derivation. Assuming that the trap state density $N_\myrm{T}(E_\myrm{T})$ is smooth around $E_\myrm{F}$, we can integrate \eqref{eq:TASchemcap} and write:
\begin{align}\label{eq:TASchemcapRes}
C_\myrm{T} &= e^2 N_\myrm{T}(E_\myrm{F}) \frac{\tilde{u}_n}{\tilde{u}_\myrm{ext}} \frac{\omega_0^2}{\omega^2 + \omega_0^2},\\
\label{eq:TASomegaRes}
\omega_{0} &=  \nu _{00}T^2  e^{-(E_\myrm{c} - E_\myrm{F})/kT}.
\end{align}

As an example, the last two equations are plotted in \figref{fig:fTAStheoRes} for the case of a discrete trap state (at $E_\myrm{c} - E_\myrm{T} = 0.3\unit{eV}$) and temperature between $150-300\unit{K}$ (blue to red). In \figrefS{fig:fTAStheoRes}{a}, the step in the capacitance at $\omega_0$ and the shift of $\omega_0$ to higher frequencies with increasing temperature are apparent. A more convenient way to represent this data is the derivative $\omega dC_\myrm{T}/d\omega$, which shows a distinct peak at the resonance frequency (\figrefS{fig:fTAStheoRes}{b}). In \figrefS{fig:fTAStheoRes}{c} we plot $\omega dC_\myrm{T}/d\omega$ in an Arrhenius-type plot in the coordinates $(1000/T, \log(\omega^{-1} T^2))$. In such a plot, $\omega_0$ reduces to a straight line from which the activation energy and the attempt frequency are determined (See \eqref{eq:TASomegaRes}).

\makefigfreetop{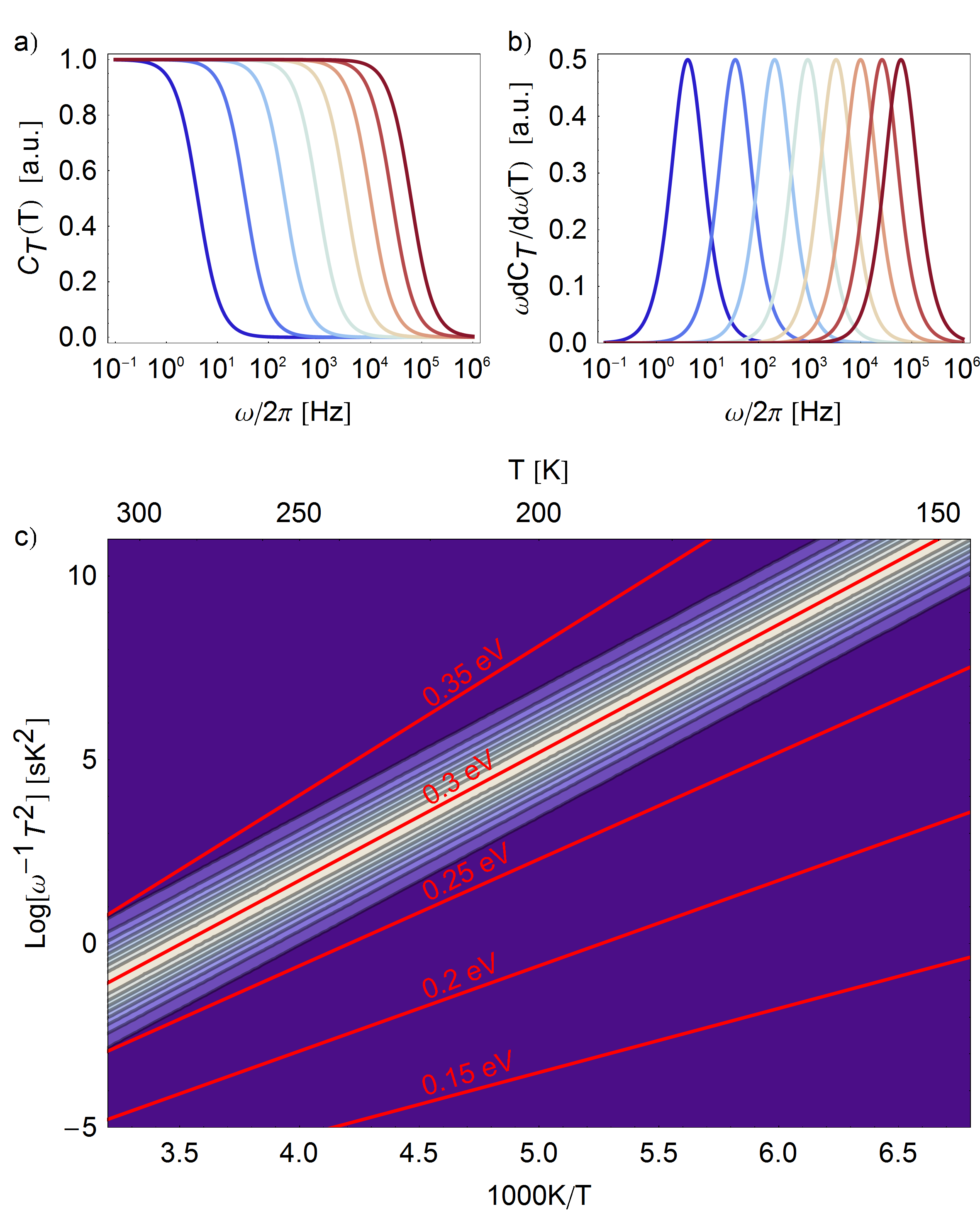}				
  {scale=0.8}		
  { %
  \mybf{a)} Simulated capacitance of a discrete trap state located $0.3\unit{eV}$ below the conduction band for temperatures between $150-300\unit{K}$. %
  \mybf{b)} Derivative $\omega dC_\myrm{T}/d\omega$ %
  \mybf{c)} Arrhenius plot of $\omega dC_\myrm{T}/d\omega$. The red lines indicate the resonance frequency ($\omega_0$) for a trap state with different energies. %
  }									
  {}												
  {fig:fTAStheoRes}									

To determine the density of trap states, we consider the magnitude of the measured capacitance. We relate the capacitance of a trap state to the real part of the capacitance of a diode ($C'(\omega)$) by integrating \eqref{eq:TASchemcapRes} over the device thickness ($d$):
\begin{align}
C'(\omega) = \int_0^d C_\myrm{T}(x,E_\myrm{F}(x))  \,dx
\end{align}
We follow \refcite{Walter1996} and use the following assumptions to perform this integration
\begin{align}
E_\myrm{T}(x) = E_\myrm{F}(x) &= E_{\myrm{F},\infty} - \frac{x}{d} e V_\myrm{bi},\\
\frac{\tilde{u}_n}{\tilde{u}_\myrm{ext}} &= \frac{x}{d}\\
\frac{\omega_0^2}{\omega^2 + \omega_0^2} &=\begin{cases}
1 \quad \omega < \omega_0\\
0 \quad \omega > \omega_0
\end{cases} \label{eq:TASassumptions}
\end{align}
where $V_\myrm{bi}$ is the built-in voltage of the diode, $E_{\myrm{F},\infty}$ is the distance between the Fermi-energy and the valence band energy at the n-type electrode. Using \eqreft{eq:TASchemcapRes}{eq:TASassumptions} the density of trap states can be expressed in terms of the measured device capacitance:
\begin{align}\label{eq:TASnt}
N_T(E_\myrm{T}) = \frac{{{V_{{\myrm{bi}}}}^2}}{{ W_\myrm{D} \left( {e{V_{{\myrm{bi}}}} - ({ E_{\myrm{F},\infty} } - E_\myrm{T})} \right)}}\frac{1}{{kT}}\frac{{{\kern 1pt} \omega \,d{C'}(\omega )}}{{d\omega }},
\end{align}
where $W_\myrm{D}$ is the width of the depletion region. $E_\myrm{T}$ is the trap energy with respect to the conduction band level and is found by inversion of \eqref{eq:TASomegaRes}:
\begin{align}\label{eq:TASet}
E_\myrm{T} = kT\log\left( 2  \nu _{00}T^2 \omega^{-1} \right)
\end{align}

\section{Experiment}\label{sec:TASExp}

\makefigfreetop{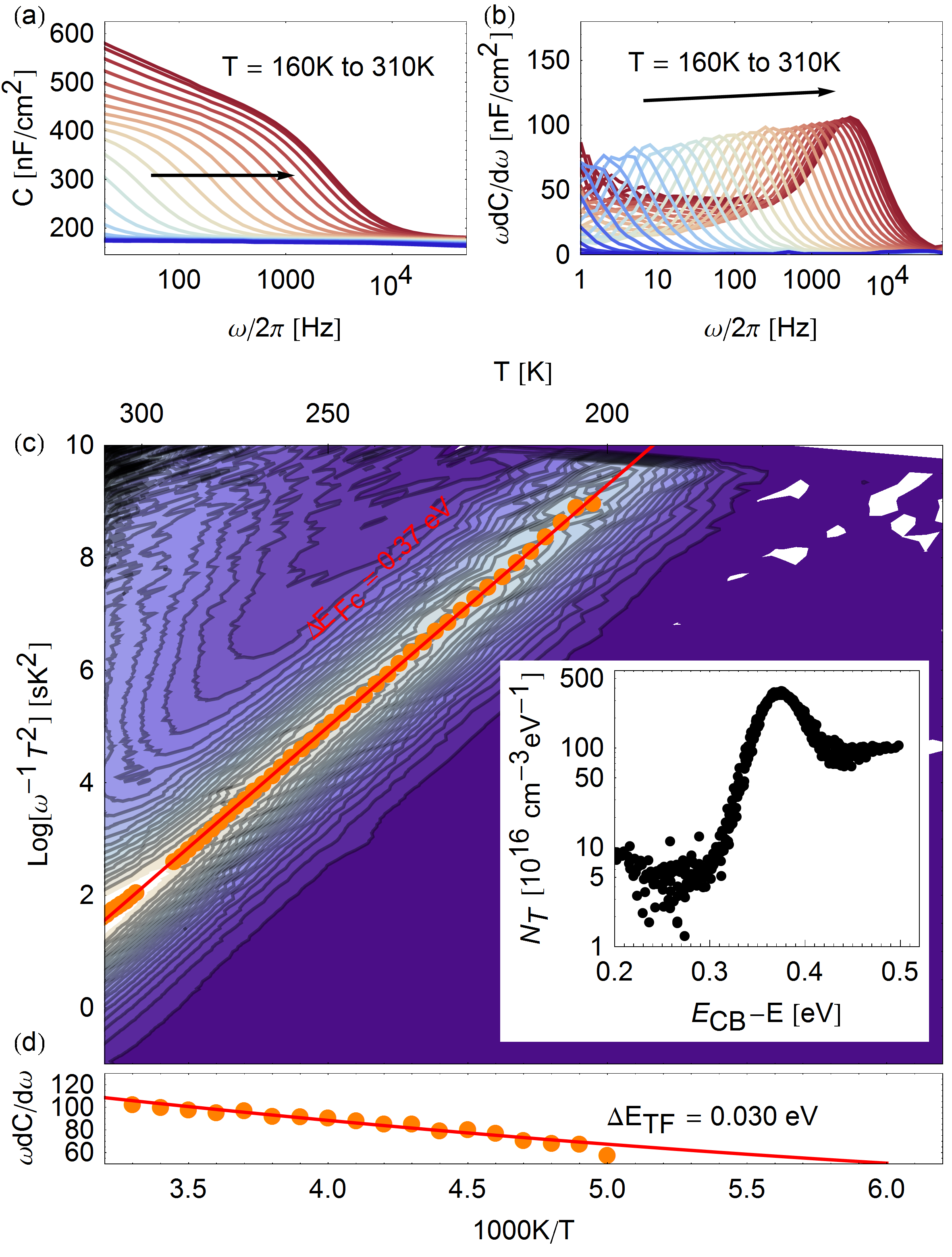}				
  {scale=0.8}		
  { %
  \mybf{a)} Real part of the device capacitance ($C'(\omega)$) for temperatures between 160~K (blue) and 310~K (red).  %
  \mybf{b)} Derivative of the real part of the capacitance ($\omega dC'/d\omega$) for same conditions as in a). %
  \mybf{c)} Arrhenius-plot of b) with extracted peak positions (orange dots) and fit of the resonance frequency (red line). %
  \mybf{d)} Fitting \eqref{eq:TAStune} to the peak data allows us to determine the detuning energy $\Delta E_\myrm{TF} = 0.03\unit{eV}$. %
  }
  {}												
  {fig:fTASexp1}									

%

In a typical TAS experiment, we measure the real part of the capacitance for frequencies between  $0.1\unit{Hz}$-$1\unit{MHz}$ using an impedance analyzer (Solartron MODULAB MTS). During the measurement, the diode is biased at $0\unit{V}$ and a modulation amplitude of $10\unit{mV}$ is applied to determine the capacitance. The capacitance measurement is performed continuously while the temperature decreases from $310\unit{K}$ to $160\unit{K}$ at a rate of $5\unit{K/min}$. Simultaneously, the temperature is measured directly at the sample and recorded for each capacitance measurement. The capacitance (${C'}(\omega)$) plotted in \figrefS{fig:fTASexp1}{a} is used to obtain the trap state density in the device. In the first step, we numerically calculate the derivative $\omega dC'(\omega )/d\omega$ from the data. In \figrefS{fig:fTASexp1}{b}, $\omega dC'(\omega )/d\omega$ is plotted  and shows the resonance peak characteristic for a discrete trap state.

We obtain the activation energy and the attempt-frequency of the trap state by fitting a line to the observed peak in the Arrhenius-type plot in \figrefS{fig:fTASexp1}{c}). We obtain ($y = 4.22 x - 11.9$), from which we calculate ${\nu _{00}} = 2.0\e{5}\unit{s^{-1} K^{-2}}$ and $E_\myrm{T} = 0.37\unit{eV}$.  We calculate the spectral trap state density by performing a coordinate transformation on the data: 
\begin{align}
\left( {\omega ,\frac{{\omega \,d{C'}(\omega )}}{{d\omega }}} \right)\underrightarrow{\text{\eqrefs{eq:TASnt}{eq:TASet}}} \left( {E,{N_T}} \right).
\end{align}
In this calculation, we use the attempt-frequency  ${\nu _{00}} = 2.0\e{5}\unit{s^{-1} K^{-2}}$ and assume a fully depleted film  ($W_\myrm{D} = d = 70{\myrm{nm}}$). In absence of a direct measurement, we estimate ${V_{{\myrm{bi}}}} = 0.6\unit{V}$ and $E_{\myrm{F},\infty} = 0.6\unit{eV}$, based on the work function difference between ITO ($\phi = 4.8\unit{eV}$) and Al ($\phi = 4.2\unit{eV}$). The final trap state density is plotted in the inset in \figrefS{fig:fTASexp1}{c}.

\section{Non-Resonant TAS}\label{sec:TASnonres}

\makefigfreetop{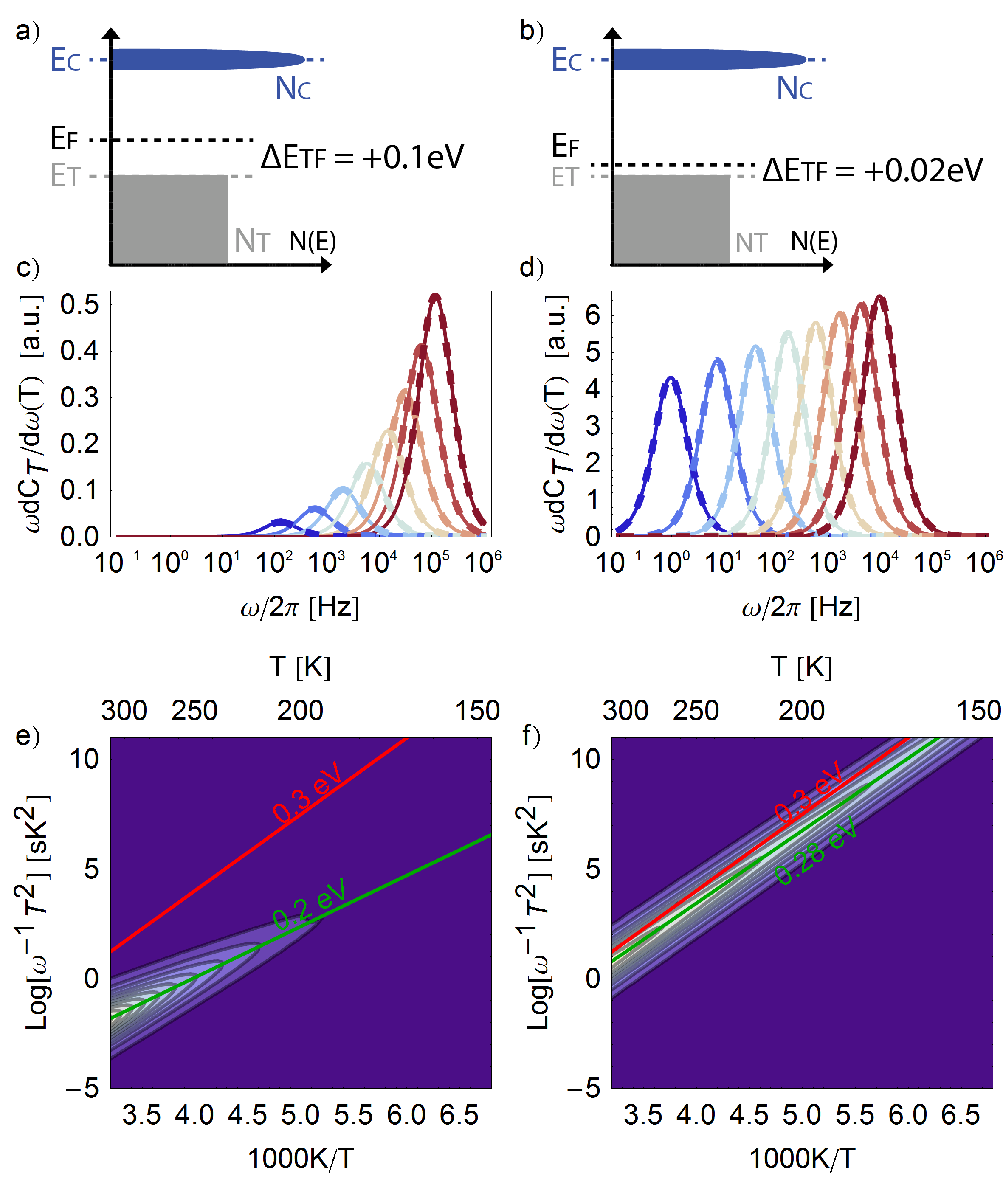}				
  {scale=0.8}		
  { %
  Band diagrams depicting non-resonant trapping where the trap state lies $0.3\unit{eV}$ below the conduction band and is \mybf{a)} $0.1\unit{eV}$ and \mybf{b)} $0.02\unit{eV}$ below the Fermi-energy.
  \mybf{c,d)} $\omega dC_\myrm{T,nr}/d\omega$ for temperatures of $150-300\unit{K}$.  %
  \mybf{e,f)} Arrhenius plot of $\omega dC_\myrm{T,nr}/d\omega$. Lines are the effective resonance frequency $\omega_{0,\myrm{nr}}$ (green) and as a comparision $\omega_0$ for the resonant case.  %
    }
  {}												
  {fig:fTAStheoOffResStep}									

In many cases the data measured on PbS NC-solids does not correspond to the ideal case shown above. Instead, the decrease of the resonance frequency with temperature is accompanied by a fast decrease of the capacitance signal. This behavior cannot be described by \eqrefs{eq:TASnt}{eq:TASet}. We therefore look for an extension of the model that can explain our experimental observations.

One of the assumptions in the preceeding derivation is that all capacitive contributions can be neglected if the trap state is not resonant with the Fermi-energy $E_\myrm{T} = E_\myrm{F}$. This assumption is good for low trap state densities, where the Fermi-level is independent of the trap states. While the Fermi-level is independent of the trap states in most polycrystalline solar cell materials, where $N_\myrm{T} = 10^{15}-10^{16} \unit{cm^{-3}}$, NC-solids can have significantly higher trap state densities ($N_\myrm{T} = 10^{17}-10^{19} \unit{cm^{-3}}$), which are able to pin the Fermi-energy. We will explain in the following how TAS can be extended to include the non-resonant capacitance contributions ($E_\myrm{T} \neq E_\myrm{F}$) and the implications for the reliable determination of trap states.

For TAS in the non-resonant case, we define the detuning between trap level and Fermi-energy $\Delta E_\myrm{TF} = E_\myrm{F} - E_\myrm{T}$. We further assume that the trap distribution is given by a step function at a characteristic trap energy $E_\myrm{T}$ as shown in \figrefS{fig:fTAStheoOffResStep}{a}. We calculate the total capacitive response in this situation by integrating \eqref{eq:TASchemcap} and obtain:
\begin{align}\label{eq:TASchemcapNonRes}
C_{\myrm{T}, \Delta E_\myrm{TF}} &=
\left[e^2 N_\myrm{T} \frac{\tilde{u}_n}{\tilde{u}_\myrm{ext}} \frac{\omega_0^2}{\omega^2 + \omega_0^2}\right] f\left(\frac{\Delta E_\myrm{TF}}{kT}\right).
\end{align}
In this integration, we have assumed that the resonance frequency does not significantly change over the integration region and is still given by \eqref{eq:TASomega0}. By assuming $\Delta E_\myrm{TF} > kT$ we rewrite \eqref{eq:TASomega0} as
\begin{align}\label{eq:TASomega0NonRes}
\omega_{0, \Delta E_\myrm{TF}} = &\nu _{00}T^2 e^{-\frac{\Delta E_\myrm{Fc}}{kT}} (1+e^{-\frac{\Delta E_\myrm{TF} }{kT} }).\\
= &\nu _{00}T^2 e^{-\frac{\Delta E_\myrm{Fc}}{kT}},
\end{align}
where $\Delta E_\myrm{Fc} = E_\myrm{c} - E_\myrm{F}$ is the distance between the conduction band and the Fermi-energy. Compared to the resonant TAS case the most important difference is that the thermal activation energy of the resonance frequency is now not related to the trap position anymore, but only to position of the Fermi-energy.

As an example we have plotted the derivative of \eqref{eq:TASchemcapNonRes} in \figref{fig:fTAStheoOffResStep} for a trap state at $0.3\unit{eV}$ below the conduction band, for two different detuning energies. In \figrefS{fig:fTAStheoOffResStep}{c} we observe that for a large detuning ($\Delta E_\myrm{TF} = 0.1\unit{eV}$) the capacitance falls off strongly when reducing the temperature from $300\unit{K}$ to $150\unit{K}$. In contrast, for a small detuning ($\Delta E_\myrm{TF} = 0.02\unit{eV}$), the capacitance is more similar to the resonant case (\figrefS{fig:fTAStheoOffResStep}{d}). Furthermore, the activation energies extracted from \figrefS{fig:fTAStheoOffResStep}{e,f} are given by the trap energy reduced by the detuning, in agreement with \eqref{eq:TASomega0NonRes}.  By comparing it to the full solution, which includes the energy dependence of $\omega_0$ in \figrefS{fig:fTAStheoOffResStep}{c,d} (dashed), we find that the assumptions leading to \eqref{eq:TASchemcapNonRes} were resonable. 

The comparison of the cases for large and small detuning shows that the measured capacitance is strongly reduced for a large detuning. If we do not take this effect into account, the trap state density determined by \eqref{eq:TASnt} will underestimate the total number of trap states by a factor of $f(\Delta E_\myrm{TF}/kT)$. This factor can be determined from the measurement data by evaluating the derivative of the capacitance at the resonance frequency:
\begin{align}\label{eq:TAStune}
\frac{\omega dC_{\myrm{T}, \Delta E_\myrm{TF}}}{d\omega} |_{\omega = \omega_{0, \Delta E_\myrm{TF}}} = -\frac{e^2 N_\myrm{T}}{2} \frac{\tilde{u}_n}{\tilde{u}_\myrm{ext}}   f\left(\frac{\Delta E_\myrm{TF}}{kT}\right).
\end{align}
The temperature dependence of this last expression is only determined by $\Delta E_\myrm{TF}$, which can be found by a fitting procedure. 

\makefig{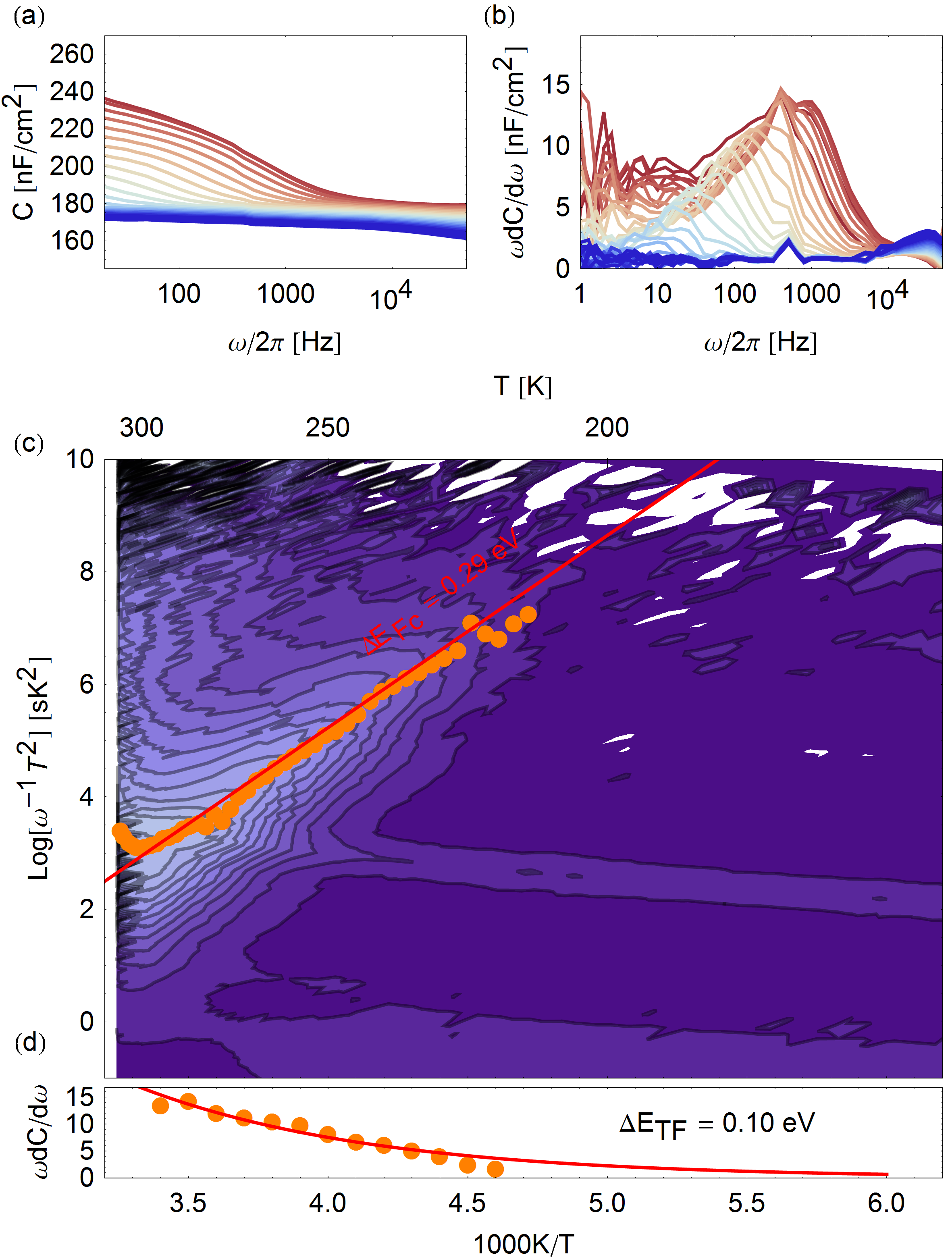}				
  {scale=0.8}		
  { %
	\mybf{a)} Real part of the device capacitance ($C'(\omega)$) of a device directly after fabrication in the temperature range $160-310\unit{K}$ (blue to red).  %
	\mybf{b)} Derivative of the real part of the capacitance ($\omega dC'/d\omega$) in a). %
	\mybf{c)} Arrhenius-plot of b) with extracted peak positions (orange dots) and fit of the resonance frequency (red line). The activation energy determined from the fit is $0.29\unit{eV}$ %
	\mybf{d)} Fitting \eqref{eq:TAStune} to the peak data allows us to determine the detuning energy $\Delta E_\myrm{TF} = 0.10\unit{eV}$. %
  }									
  {}												
  {fig:fTASexpNonRes}									

%

In TAS measurements on freshly fabricated PbS NC-based devices, we usually observe a non-resonant TAS response as shown in \figref{fig:fTASexpNonRes}. With decreasing temperature, we observe a quickly receeding capacitance, which is characteristic for the non-resonant TAS response. The activation energy is determined from \figrefS{fig:fTASexpNonRes}{c} to be $\Delta E_\myrm{Fc} = 0.29\unit{eV}$. The detuning is determined by the fit of \eqref{eq:TAStune} in \figrefS{fig:fTASexpNonRes}{d} and yields $\Delta E_\myrm{TF} = 0.10\unit{eV}$. Adding these two numbers gives the real trap level position at $0.39\unit{eV}$ below the conduction band. The large detuning further implies that the total trap density of $3\e{16}\unit{cm^{-3}}$ in this measurement underestimates the real trap density by a factor of 50 (at $300\unit{K}$). The real trap density in this measurement is therefore on the order of $1.5\e{18}\unit{cm^{-3}}$.

We compare this to the measurement shown in \figref{fig:fTASexp1}, which was performed on the same device after two days in air. We determine $\Delta E_\myrm{Fc} = 0.37\unit{eV}$  and $\Delta E_\myrm{TF} = 0.03\unit{eV}$. The real trap level position is thus $0.40\unit{eV}$, which is in excellent agreement with the measurement on the fresh device ($0.39\unit{eV}$). We conclude that air exposure does not change the energetic position of the trap state but rather shifts the Fermi-level down by $0.07\unit{eV}$. This is consistent with our previous findings from a combination of Fourier Transform Photocurrent Spectroscopy (FTPS) and Deep Level Transient Spectroscopy (DLTS) \refcite{Bozyigit2013b}.

\section{Conclusion}

We have extended the theory of TAS to account for situations where the Fermi level is pinned by large density of trap states.  Our approach enables us to correctly quantify the density of trap states and determine the position of the Fermi level relative to the trap states in the semiconductor under investigation.   We show that this approach is particularly applicable to NC solids and promises to be of use in developing a better understanding of the interplay between device fabrication and trap states in solution processed semiconductors.



\bibliography{./Bib/thesis}

\end{document}